\documentclass[11pt]{article}
\usepackage{graphics}
\usepackage{theorem}
\usepackage{latexsym}
\usepackage{amsfonts}
\usepackage{amssymb}
\usepackage{amsmath}
\usepackage{natbib}
\usepackage{url}
\theoremstyle{break}

\newtheorem{Remark}{Remark}[section]
\newtheorem{Assumption}{Assumption}[section]
\newtheorem{Proposition}{Proposition}[section]
\newtheorem{Example}{Example}[section]

\numberwithin{equation}{section}
\title{On the Informativeness of Specification Tests for Estimator Validity\footnote{Previous versions of the manuscript were circulated under the title ``A Misuse of Specification Tests." I am grateful to Ryo Okui and Wenjie Wang for their valuable comments. I also thank seminar participants at Keio University, Kyoto University, and the University of Tokyo for their helpful feedback. This research was supported by JSPS KAKENHI Grant Number 21K01427. 
}}
\author{Naoya Sueishi \\
    {\it Kobe University\footnote{Graduate School of Economics, 2-1 Rokkodai-cho, Nada-ku, Kobe, Hyogo 657-8501, Japan. Email: sueishi@econ.kobe-u.ac.jp. }}
}
\date{\today}

\begin{document}
\maketitle

\abstract{
	Empirical researchers often use model specification tests, such as Hausman tests and overidentifying restrictions tests, to assess the validity of estimators rather than the correctness of models. This paper examines the extent to which such tests are informative about the presence of asymptotic bias in estimators. Under a local misspecification framework, we show that the directions of local deviation from a benchmark distribution to which locally unbiased specification tests have nontrivial power are orthogonal to those that induce asymptotic bias in asymptotically efficient estimators. Consequently, specification tests generally provide limited information about estimator validity unless researchers impose additional untestable assumptions on the form of misspecification. We further demonstrate that, when used as estimator selection rules, Hausman tests can lead to the choice of inefficient estimators that exhibit asymptotic bias, even when efficient estimators remain asymptotically unbiased. These findings highlight fundamental limitations in using specification tests to guide estimator choice.
}
\\


\newpage

\section{Introduction}

Model specification tests, such as Hausman tests (\citealt{hausman1978specification}) and overidentifying restrictions tests (\citealt{sargan1958estimation};  \citealt{hansen1982large}), are widely used in empirical studies, although finding a true model is rarely the primary goal.
Empirical researchers often conduct these tests to guide the choice of estimator.
For example, the Durbin-Wu-Hausman (DWH) test (\citealt{durbin1954}; \citealt{wu1973alternative}) is typically used to decide between the ordinary least squares (OLS) estimator and the two-stage least squares (2SLS) estimator.
Such practices continue to be prevalent in empirical applications, despite a substantial literature questioning the usefulness of pretests in terms of potential distortions in subsequent inference (see, e.g., \citealt{guggenberger2010impact} and \citealt{guggenberger2010joe}).

This paper examines the effectiveness of specification tests in assessing the asymptotic validity of estimators.
It is commonly believed that failure to reject the null hypothesis provides evidence that an estimator based on the null model is reliable, whereas rejection suggests that the estimator may suffer from misspecification bias.
We investigate the extent to which specification tests can inform us about the presence of asymptotic bias in the associated estimators.

Our first contribution is to show that neither rejection nor failure to reject the null hypotheses of standard specification tests can generally be interpreted as evidence about the asymptotic validity of estimators unless one is willing to impose untestable assumptions.
We analyze how the asymptotic bias of asymptotically efficient estimators and the power of specification tests respond to local deviations from a benchmark distribution. Using the framework of \cite{chen2018overidentification}, we formalize these deviations using a score function of a path, which can be decomposed into two orthogonal components. We show that the asymptotic bias of efficient estimators depends only on one component, whereas the local power of locally unbiased tests depends only on the other.  
Although similar relationships between estimators and specification test statistics have been observed in some special cases, such as the relationship between the efficient GMM estimator and the $J$-test statistic (e.g., \citealt{hall2005}), our analysis shows that the phenomenon is not specific to particular estimators or test statistics. Rather, it reflects a general structural property of overidentified semiparametric models.

Our second contribution is to show that Hausman tests can be misleading when interpreted as direct evidence about estimator validity. In particular, a Hausman test may reject its null hypothesis even when the efficient estimator used to construct the test statistic remains asymptotically unbiased, while the inefficient estimator is asymptotically biased. This may seem paradoxical, because Hausman tests are often interpreted as testing the validity of the efficient estimator under the maintained assumption that the inefficient estimator is valid. We show that this apparent inconsistency arises from the way the null and alternative hypotheses are formulated, an issue that is often left implicit in informal discussions of Hausman tests. We specify null and alternative hypotheses under which Hausman tests are locally unbiased. Furthermore, we show that Hausman tests may lead to the selection of estimators with inferior asymptotic properties unless one imposes untestable a priori assumptions regarding the directions of local deviation.

The present paper differs from the existing literature on pretesting in its main focus.
Many studies have investigated the impact of pretesting or model selection on subsequent inference.
Examples include \cite{jud_boc1978}, \cite{potscher1991effects}, \cite{kabaila1995effect}, \cite{leeb2005model, leeb2006can}, and \cite{andrews2009hybrid, andrews2009incorrect}.
\cite{guggenberger2010impact,guggenberger2010joe}, \cite{guggenberger2012size}, and \cite{dok_wan2021} specifically investigated the impact of specification tests on subsequent inference. \cite{guggenberger2012size} also provide several examples of empirical studies in which overidentifying restrictions tests are used as a pretest.
These studies show that inference ignoring the effect of pretesting can be highly misleading, with a particular emphasis on size distortion in second-stage tests. In contrast, this paper questions the very rationale for pretesting itself by uncovering its structural limitations. Indeed, our orthogonal decomposition also provides an explanation for the size distortions documented by \cite{guggenberger2012size}.

The local asymptotic framework is also widely used in the context of robust estimation and inference, particularly for models specified by moment restrictions.
\cite{kitamura2013robustness} proposed a robust point estimator when the data are generated by a local deviation from a benchmark distribution.
\cite{armstrong2021sensitivity} proposed confidence intervals that take into account the potential bias resulting from a local deviation.
See also \cite{andrews2017measuring}, \cite{bonhomme2022minimizing} and \cite{andrews2025purpose} for related issues.
Our results suggest that, when misspecification is a concern, it may be preferable to adopt robust methods from the outset rather than relying on specification pretests.

The remainder of the paper is organized as follows.
Section~\ref{sec:setup} introduces the setting of our analysis, and Section~\ref{sec:main} presents the main results of the paper.
Sections~\ref{sec:jtest} and \ref{sec:dwh} show how the results of Section~\ref{sec:main} apply to popular models in econometrics.
Section~\ref{sec:jtest} mainly illustrates how existing results on the efficient GMM estimator and the $J$-test fit into our general framework.
In contrast, Section~\ref{sec:dwh} provides new implications for the DWH test under local misspecification.
Section~\ref{sec:simulation} presents Monte Carlo studies that illustrate our theoretical findings, and Section~\ref{sec:conclusion} concludes.
The Appendix derives the tangent set for models defined by conditional moment restrictions, which is used in Section~\ref{sec:dwh}.

\section{Setup}\label{sec:setup}

Let $P$ be a probability distribution defined on a sample space $\mathcal{X}$, and let $\mathcal{M}$ denote the set of all distributions on $\mathcal{X}$. 
The distribution $P$ may be motivated by an underlying economic model or regarded as an idealized distribution from which researchers hope to draw a sample. 
We assume that $P$ belongs to a semiparametric model $\mathbf{P}$, which is represented as a set of distributions on $\mathcal{X}$. 
The finite-dimensional parameter of interest, $\theta_0 \in \Theta$, is defined as $\theta_0 = \psi(P)$ for a smooth functional $\psi: \mathbf{P} \to \Theta$. 
This parameter typically corresponds to a component of a structural model that captures the economic mechanism of interest.
Although we focus on $\theta_0$, the model may additionally involve infinite-dimensional nuisance parameters.

A random sample $\{X_1, \dots, X_n\}$ is drawn from a distribution $\mu$ on $\mathcal{X}$.
Let $\hat{\theta}_n:\{X_i\}_{i=1}^n \to \Theta$ be an asymptotically efficient estimator for $\theta_0$ at $P$ relative to $\mathbf{P}$, meaning that it attains the smallest asymptotic variance matrix among all regular estimators. We consider the possibility that the data generating process (DGP) $\mu$ deviates locally from $P$, which may arise from data contamination, measurement error, model misspecification, or other sources.
If the deviation vanishes at the rate $n^{-1/2}$, then $\hat{\theta}_n$ remains consistent for $\theta_0$, but may be asymptotically biased in the sense that the asymptotic mean of $\sqrt{n}(\hat{\theta}_n - \theta_0)$ is nonzero.
In what follows, we say that an estimator is valid if its asymptotic mean is zero.

To detect the local deviation from $P$, we consider a specification test $\phi_n: \{X_i \}_{i=1}^n \to [0,1]$ for the following null and alternative hypotheses:
\begin{equation}\label{hypotheses1}
	H_0: \mu \in \mathbf{P} \quad \text{vs.} \quad H_1: \mu \in \mathcal{M} \setminus \mathbf{P}.
\end{equation}
Whether the test exhibits nontrivial power against the local deviation depends on the direction of the deviation from $P$.

We investigate the relationship between the estimator and the test.
Our concern is whether the test can detect local deviations that induce asymptotic bias in the estimator.
As we shall see shortly, not all local deviations lead to asymptotic bias in $\hat{\theta}_n$.
Accordingly, if the test is used to assess the validity of the estimator, it would be desirable for the test to have nontrivial local power if and only if the estimator is asymptotically biased.
The asymptotic behavior of $\hat{\theta}_n$ and $\phi_n$ depends on the direction of the local deviation.
To formalize this notion, we adopt the framework of \cite{chen2018overidentification}.

The direction of a local deviation from $P$ is defined in terms of the score function of a path.
A path $t \mapsto P_t$ is a function defined on $[0,\epsilon)$ for some $\epsilon >0$ such that $P_t \in \mathcal{M}$ for all $t \in [0,\epsilon)$ and $P_0 = P$.
In other words, a path is a parametric model indexed by a scalar parameter $t$, with the ``true'' value of $t$ being zero.
We focus on paths $t \mapsto P_{t,g} \in \mathcal{M}$ that satisfy
\begin{equation}\label{hellinger}
	\lim_{t \to 0} \int  \left(\frac{dP_{t,g}^{1/2} - dP^{1/2}}{t} - \frac{1}{2} g dP^{1/2} \right)^2 = 0
\end{equation}
for some function $g : \mathcal{X} \to \mathbb{R}$. 
If this holds, the path is said to be differentiable in quadratic mean at $t=0$.
The function $g$ is referred to as the score function because it is typically given by 
\[
	g(x) = \left. \frac{\partial}{\partial t} \log dP_{t,g}(x)\right|_{t=0}.
\]
If a function $g$ that satisfies \eqref{hellinger} exists, it must satisfy $\mathbb{E}[g(X)]=0$ and $\mathbb{E}[g^2(X)]<\infty$, where $\mathbb{E}$ denotes the expectation with respect to $P$.
Thus, the set of all possible score functions is given by
\[
	L_0^2(P) \equiv \left\{ g: \mathcal{X} \to \mathbb{R}: \mathbb{E}[g(X)]=0 \ \text{and} \ \mathbb{E}[g^2(X)]<\infty \right\}.
\]

The directions of local deviation that are consistent with the null hypothesis are captured by the tangent set.
We define
\[
	T(P)= \left\{ g \in L_0^2(P) : \eqref{hellinger} \ \text{holds for some} \ t \mapsto P_{t,g} \in \mathbf{P} \right\},
\]
which is called the tangent set of model $\mathbf{P}$ at $P$.
Intuitively, it consists of all directions along which the distribution can locally deviate from $P$ while remaining within the model $\mathbf{P}$.
Throughout the paper, we impose the following condition on $T(P)$:
\begin{Assumption}\label{linear_space}
	$T(P)$ is a linear space.
\end{Assumption}
This assumption is satisfied in many semiparametric models.
The examples considered in Sections~\ref{sec:jtest} and \ref{sec:dwh} also satisfy Assumption~\ref{linear_space}.

Any score $g \in L_0^2(P)$ can be decomposed into two orthogonal components under Assumption~\ref{linear_space}.
Let $\bar{T}(P)$ be the closure of $T(P)$ under $\| \cdot \|_{P,2}$, where $\|g \|_{P,2}^2=\mathbb{E}[g^2(X)]$ for $g \in L_0^2(P)$. We define its orthogonal complement as
\[
	\bar{T}(P)^\bot = \left\{ f \in L_0^2(P): \mathbb{E}[f(X)g(X)] =0 \ \text{for all} \ g \in \bar{T}(P)   \right\}.
\]
Then, by the orthogonal projection theorem for Hilbert spaces, we obtain the orthogonal decomposition $L_0^2(P) = \bar{T}(P) \oplus \bar{T}(P)^\bot$.
Consequently, any score $g$ can be decomposed as $g = \Pi_{T}(g) + \Pi_{T^\bot}(g)$, where $\Pi_T$ and $\Pi_{T^\bot}$ denote the projection with respect to $\| \cdot \|_{P,2}$ onto $\bar{T}(P)$ and $\bar{T}(P)^\bot$, respectively.

\cite{chen2018overidentification} showed that local overidentification is both necessary and sufficient for the existence of an asymptotically efficient estimator for $\theta_0$ and a locally unbiased test for \eqref{hypotheses1}.
They say that $P$ is locally overidentified by $\mathbf{P}$ if $\bar{T}(P) \neq L_0^2(P)$.
This definition generalizes the classical definition of overidentification, which compares the number of moments with the number of parameters.
See also Theorem 2.1 of \cite{newey1994asymptotic} for a related issue.
Intuitively, if $\bar{T}(P) = L_0^2(P)$, every local deviation from $P$ is compatible with the model, so the model imposes no effective local restriction on $P$.

Building on the framework above, we investigate the relationship between the asymptotically efficient estimator and the locally unbiased test. Specifically, we consider the case where the random sample is drawn from $P_{1/\sqrt{n},g}$ for some $g \in L_0^2(P)$, so that the DGP represents a local deviation from $P$. We then analyze how the asymptotic bias of the estimator and the local power of the test depend on $g$.

To facilitate understanding of the framework above, we illustrate how the $J$-test fits within it through the following example.

\begin{Example}

	The parameter of interest $\theta_0 \in \Theta \subset \mathbb{R}^p$ is the unique vector that satisfies the moment restrictions:
	\begin{equation}\label{iden_asm}
		0 = \mathbb{E}[m_{\theta_0}(X)] =  \int m_{\theta_0} dP,
	\end{equation}
	where $m: \mathcal{X} \times \Theta \to \mathbb{R}^l$ is a known vector-valued function with $l >p$.
	The model is semiparametric in the sense that $P$ is an infinite-dimensional nuisance parameter.
	If a random sample is drawn from $\mu$, then the null hypothesis of the $J$-test is
	\[
		H_0: \int m_\theta d\mu =0 \ \text{for some} \ \theta \in \Theta.
	\]
	Equivalently, the null hypothesis can be written as $\mu \in \mathbf{P}$ where
	\begin{equation}\label{Pset}
		\mathbf{P}= \left\{ Q \in \mathcal{M}: \int m_\theta dQ=0 \ \text{for some} \ \theta \in \Theta \right\}.
	\end{equation}
	That is, $\mathbf{P}$ is the set of all distributions on $\mathcal{X}$ that satisfy the moment restrictions for some $\theta \in \Theta$.
\end{Example}

There are many possible ways to define a path.
For instance, if $1+tg$ is nonnegative for sufficiently small $t$, let $P_{t, g}= (1 +tg) P$ for some $g \in L_0^2(P)$. Then, it satisfies $P_{0,g} =P$ and $\int dP_{t,g} =1$. Moreover, we have
\[
	\left. \frac{\partial}{\partial t} \log dP_{t,g}(x)\right|_{t=0} = g.
\]
Thus, $P_{t,g}$ is a path whose score function is $g$.
In what follows, the specific choice of the path is irrelevant; what matters is solely the set of score functions it generates.
In Section~\ref{sec:jtest}, we present an alternative formulation of \eqref{Pset} and introduce the tangent set of $\mathbf{P}$ at $P$.

\begin{Remark}
	The $J$-test is often used to test the following null hypothesis:
	\begin{equation}\label{jnull}
		H_0: \mathbb{E}[m_{\theta_0}(X)]=0.
	\end{equation}
	A problem of testing \eqref{jnull} has been pointed out by some studies  (see \citealt{deaton2010instruments} and \citealt{parente2012cautionary} among others). 
	The point is that \eqref{iden_asm} is the identifying assumption for $\theta_0$ that must be imposed before the analysis.
	The $J$-test does not verify whether \eqref{jnull} holds because it has no power when $\int m_\theta d\mu =0$ holds for some $\theta \in \Theta$. 
	
\end{Remark}

\begin{Remark}
	The local misspecification framework is commonly used to analyze the properties of estimators and test statistics under misspecification (\citealt{newey1985generalized}).  
	In this setting, we do not literally believe that the DGP approaches $P$ as the sample size increases; rather, the framework serves as a mathematical device to capture small deviations from $P$.  
	A similar idea is also employed in the literature on weak instruments.  
	We adopt this framework because we believe that it provides a reasonable approximation to reality.
\end{Remark}

\begin{Remark}
	\cite{andrews2025purpose} distinguish between econometric misspecification and statistical misspecification. The distinction is important when the parameter of interest can be defined independently of a statistical model. Since the parameter of interest in this paper is the model parameter $\theta_0$, we consider only statistical misspecification. While analyses of misspecified models often consider pseudo-true parameters, such parameters are not the object of our interest.
\end{Remark}

\section{Main Results} \label{sec:main}

This section shows that the relationship between specification tests and asymptotically efficient estimators is determined by a general structural property of semiparametric models that locally overidentify $P$.
In particular, the directions of local deviation detectable by locally unbiased specification tests are orthogonal to those that induce asymptotic bias in asymptotically efficient estimators.
As a result, specification tests provide inherently limited information about the asymptotic bias of efficient estimators.
We also show that Hausman tests may lead researchers to select inefficient estimators with asymptotic bias, even when efficient estimators remain asymptotically unbiased.

\subsection{Orthogonality}\label{subsec:orthogonality}

To investigate the properties of specification tests, we introduce some additional definitions from \cite{chen2018overidentification}.
A test $\phi_n$ for \eqref{hypotheses1} has local asymptotic level $\alpha$ if 
\[
	\limsup_{n \to \infty} \int \phi_n dP_{1/\sqrt{n}, g}^n  \le \alpha  \quad 
\]
for any path $t \mapsto P_{t,g} \in \mathbf{P}$, where $P_{1/\sqrt{n},g}^n$ denotes the joint distribution of $\{X_1, \dots, X_n\}$.
A test $\phi_n$ has a local asymptotic power function $\pi: L_0^2(P) \to [0,1]$ if
\[
	\lim_{n \to \infty} \int \phi_n dP_{1/\sqrt{n},g}^n = \pi(g)
\]
for any path $t \mapsto P_{t, g} \in \mathcal{M}$.
Finally, the test is locally unbiased if it satisfies $\pi(g) \le \alpha$ for all $t \mapsto P_{t,g} \in \mathbf{P}$ and $\pi(g) \ge \alpha$ for all $t \mapsto P_{t,g} \in \mathcal{M} \setminus \mathbf{P}$.
These definitions are stated in terms of paths, but the local asymptotic power function depends only on the score generated by the path and not on other features of the path (\citealt{chen2018overidentification}).

It follows from the above definitions that locally unbiased tests do not have nontrivial local power in directions in $T(P)$. Since $\pi$ depends continuously on the score under $\| \cdot \|_{P,2}$, the same conclusion extends to directions in $\bar{T}(P)$.
Consequently, locally unbiased tests can have nontrivial local power only through the component $\Pi_{T^\bot}(g)$.

For standard locally unbiased specification tests, this restriction is typically reflected in the fact that the test statistics are asymptotically constructed from elements of $\bar{T}(P)^\bot$.
To see this, consider a specification test statistic that is asymptotically chi-square distributed under $P$ and admits the expansion
\[
	T_n = \sum_{j=1}^K \left( \frac{1}{\sqrt{n}} \sum_{i=1}^n f_j(X_i) \right)^2 +o_P(1),
\]
where $f_1, \dots, f_K$ are orthonormal and $K$ determines the degrees of freedom.
Since the path is differentiable in quadratic mean at $t=0$, $P_{t,g}$ satisfies
\begin{equation}\label{lan}
	\log \prod_{i=1}^n \frac{dP_{1/\sqrt{n}, g}}{dP}(X_i) = \frac{1}{\sqrt{n}} \sum_{i=1}^n g(X_i) - \frac{1}{2} \| g \|_{P,2}^2 + o_P(1)
\end{equation}
for all $g \in L_0^2(P)$ (Lemma 25.14 of \citealt{van1998asymptotic}).
The advantage of having \eqref{lan} hold is that it allows us to derive the asymptotic distribution of a statistic under $P_{1/\sqrt{n},g}$ using the asymptotic distribution under $P$.
Let $f=(f_1,\dots,f_K)'$.
By Le Cam's third lemma,
\[
	\frac{1}{\sqrt{n}} \sum_{i=1}^n f(X_i) \stackrel{g}{\rightsquigarrow} N \left(\mathbb{E}[f(X)g(X)], I \right),
\]
where $\stackrel{g}{\rightsquigarrow}$ denotes the weak convergence under $P_{1/\sqrt{n}, g}$.
The continuous mapping theorem then yields
\[
	T_n \stackrel{g}{\rightsquigarrow} \chi_K^2 \left(\sum_{j=1}^K \mathbb{E}[f_j(X)g(X)]^2 \right),
\]
where $\chi_k^2(a)$ denotes the noncentral chi-square distribution with degrees of freedom $k$ and the noncentrality parameter $a$.
For such tests, local unbiasedness requires $f_j \in \bar{T}(P)^\bot$ for all $j=1, \dots, K$.

Next, we investigate the asymptotic bias of asymptotically efficient estimators.
We impose the following condition on $\theta_0=\psi(P)$:

\begin{Assumption}\label{regular_parameter}
	The functional $\psi$ is pathwise differentiable at $P$ with respect to $T(P)$ in the sense that there exists $\nu$ such that $\mathbb{E}[\nu(X)\nu(X)']<\infty$ and 
	\[
		\lim_{t \to 0} \frac{\psi(P_{t,g})-\psi(P)}{t} = \mathbb{E}[\nu(X) g(X)]
	\]
	for any $g \in T(P)$.
\end{Assumption}
Under Assumption~\ref{regular_parameter}, any asymptotically efficient estimator for $\theta_0$ satisfies
\[
	\sqrt{n}(\hat{\theta}_n - \theta_0) = \frac{1}{\sqrt{n}} \sum_{i=1}^n \nu(X_i) + o_P(1).
\]
The function $\nu$ is the efficient influence function whose components belong to $\bar{T}(P)$.
Assumption~\ref{regular_parameter} is satisfied if there is a linear regular estimator for $\psi(P)$ (\citealt{van1991differentiable}).

By Le Cam's third lemma, we obtain
\[
	\sqrt{n}(\hat{\theta}_n - \theta_0) \stackrel{g}{\rightsquigarrow} N(\mathbb{E}[\nu(X)g(X)], \mathbb{E}[\nu(X) \nu(X)']).
\]
This means that the asymptotic distribution of $\hat{\theta}_n$ is unaffected by the local deviation if $\nu(X)$ is uncorrelated with $g(X)$.
Since each component of the efficient influence function belongs to $\bar{T}(P)$, the asymptotic bias depends only on $\Pi_{T}(g)$.
If $g \in \bar{T}(P)^\bot$, then the asymptotic distribution of $\hat{\theta}_n$ under $P_{1/\sqrt{n}, g}$ is the same as that under $P$.

Combining these results, we obtain the following proposition.

\begin{Proposition} \label{orthogonality}
	Suppose that $P$ is locally overidentified by $\mathbf{P}$ and that Assumptions~\ref{linear_space} and \ref{regular_parameter} hold.
	Let $\hat{\theta}_n$ be an asymptotically efficient estimator for $\theta_0$ at $P$, and let $\phi_n$ be a locally unbiased asymptotic level-$\alpha$ test for (\ref{hypotheses1}) whose local asymptotic power function is $\pi$.
	Then, $\hat{\theta}_n$ can be asymptotically biased only if $\Pi_T(g) \neq 0$.
	Moreover, $\pi(g)>\alpha$ only if $\Pi_{T^\bot}(g) \neq 0$.
\end{Proposition}

Proposition~\ref{orthogonality} states that the directions of local deviation detectable by locally unbiased specification tests are orthogonal to those that generate asymptotic bias in asymptotically efficient estimators.
Specification tests are of limited value for assessing the validity of estimators, since they detect only the component of deviation that does not generate asymptotic bias and fail to detect the component that does.
While such orthogonality has been recognized in special cases such as the relationship between the efficient GMM estimator and the $J$-test, our result shows that it is in fact a general feature of the local geometry of models that locally overidentify $P$, independent of any particular estimator or test statistic.

Proposition~\ref{orthogonality} does not necessarily imply that specification pretests are entirely uninformative.
One might intuitively believe that $\Pi_T(g)$ tends to be zero when $\Pi_{T^\bot}(g)=0$ and tends to be nonzero when $\Pi_{T^\bot}(g)\neq 0$.
Under such circumstances, specification tests would indeed be informative about the presence or absence of asymptotic bias.
Although this line of reasoning may be plausible, it should be recognized that it relies on beliefs that cannot be verified from the data.
Moreover, even if such a belief can be justified, specification tests still provide no information about the magnitude of the bias. In particular, the magnitude of the test statistic provides no guidance as to whether the estimator is more or less severely biased.

\subsection{Maintained Hypothesis and Hausman Test} \label{subsec:hausman}

We now introduce the null and alternative hypotheses under a maintained hypothesis:
\begin{equation}\label{hypotheses2}
	H_0: \mu \in \mathbf{P} \quad \text{vs.} \quad H_1: \mu \in \mathbf{M} \setminus \mathbf{P},
\end{equation}
where $\mathbf{M}$ is another semiparametric model such that $\mathbf{P} \subset \mathbf{M} \subset \mathcal{M}$. That is, we assume $\mu \in \mathbf{M}$ under both the null and alternative hypotheses. The parameter of interest $\theta_0$ is defined by $\theta_0 = \psi(P) = \varphi(P)$ for pathwise differentiable functionals $\psi:\mathbf{P} \to \Theta$ and $\varphi:\mathbf{M} \to \Theta$ such that $\psi(Q) = \varphi(Q)$ for all $Q \in \mathbf{P}$.

In this setting, we can decompose $L_0^2(P)$ using the tangent set of model $\mathbf{M}$ at $P$.
Define 
\[
	M(P) = \left\{g \in L_0^2(P): \eqref{hellinger} \ \text{holds for some} \ t \mapsto P_{t,g} \in \mathbf{M} \right\}
\]
and assume that $M(P)$ is linear.
Let $\bar{M}(P)$ be the closure of $M(P)$ under $\| \cdot \|_{P,2}$.
Since $\bar{T}(P) \subset \bar{M}(P)$, we have
\[
	L_0^2(P) = \bar{T}(P) \oplus \left\{ \bar{T}(P)^\bot \cap \bar{M}(P) \right\} \oplus \bar{M}(P)^\bot.
\]
Thus, any score $g$ can be decomposed as $g  = \Pi_T(g) + \Pi_{T^\bot \cap M}(g) + \Pi_{M^\bot}(g)$.
The same decomposition is also discussed in \cite{chen2018overidentification}.

It is clear that locally unbiased tests for \eqref{hypotheses2} can have nontrivial local power for the deviation $P_{1/\sqrt{n}, g}$ only if $\Pi_{T^\bot \cap M}(g) \neq 0$.
In contrast, asymptotically efficient estimators for $\theta_0$ in $\mathbf{P}$ can be asymptotically biased only if $\Pi_T(g) \neq 0$.
Therefore, once again, specification tests alone cannot detect local deviations that cause asymptotic bias in efficient estimators.

The Hausman test is a representative example of tests that fall within the above setup.
Let $\hat{\theta}_n$ and $\tilde{\theta}_n$ be asymptotically efficient estimators for $\theta_0$ in models $\mathbf{P}$ and $\mathbf{M}$, respectively.
Since $\mathbf{P} \subset \mathbf{M}$, $\hat{\theta}_n$ is more efficient than $\tilde{\theta}_n$.
Let $V$ denote the asymptotic variance matrix of $\tilde{\theta}_n -\hat{\theta}_n$, and let $\hat{V}$ be a consistent estimator of $V$.
Then, the test statistic is given by
\begin{equation}\label{hausman_stat}
	T_n = n(\tilde{\theta}_n - \hat{\theta}_n)' \hat{V}^{-1} (\tilde{\theta}_n -\hat{\theta}_n),
\end{equation}
where $\hat{V}^{-1}$ is replaced with the Moore-Penrose generalized inverse if $\hat{V}$ is singular.

To analyze the local properties of the Hausman test, we write the two estimators in their asymptotically linear form:
\begin{align*}
	 & \sqrt{n}(\hat{\theta}_n - \theta_0) = \frac{1}{\sqrt{n}} \sum_{i=1}^n \nu(X_i) + o_P(1),    \\
	 & \sqrt{n}(\tilde{\theta}_n - \theta_0) = \frac{1}{\sqrt{n}} \sum_{i=1}^n \tau(X_i) + o_P(1)
\end{align*}
for some $\nu \in \bar{T}(P)$ and $\tau \in \bar{M}(P)$. Here, $\nu \in \bar{T}(P)$ means that each component of $\nu$ belongs to $\bar{T}(P)$, and similarly for $\tau \in \bar{M}(P)$.
It is well known that the efficient influence function is obtained by projecting any other influence function onto the tangent space (see, e.g., \citealt{van1998asymptotic}).
Since $\hat{\theta}_n$ is more efficient than $\tilde{\theta}_n$, we have $\nu =\Pi_T(\tau)$ and $\tau - \nu \in \bar{T}(P)^\bot \cap \bar{M}(P)$.
This implies the well-known fact that the asymptotic variance of $\tilde{\theta}_n - \hat{\theta}_n$ equals the difference between the asymptotic variances of $\tilde{\theta}_n$ and $\hat{\theta}_n$.
Moreover, since $\tau -\nu \in \bar{T}(P)^\bot \cap \bar{M}(P)$, there exist orthonormal functions $f_1, \dots, f_K \in \bar{T}(P)^\bot \cap \bar{M}(P)$ such that
\[
	T_n = \sum_{j=1}^K \left( \frac{1}{\sqrt{n}} \sum_{i=1}^n f_j(X_i)  \right)^2 +o_P(1),
\]
where $K$ is the rank of the variance matrix of $\tau(X)-\nu(X)$. 
Thus, we have
\[
	T_n \stackrel{g}{\rightsquigarrow} \chi_K^2 \left(\sum_{j=1}^K \mathbb{E}[f_j(X)g(X)]^2 \right).
\]
This shows that the Hausman test is locally unbiased for \eqref{hypotheses2}.
The test has nontrivial local power only through $\Pi_{T^\bot \cap M}(g)$.

Next, we investigate the asymptotic properties of two estimators. 
Applying Le Cam's third lemma, we obtain
\begin{align*}
	 & \sqrt{n}(\hat{\theta}_n - \theta_0) \stackrel{g}{\rightsquigarrow} N(\mathbb{E}[\nu(X) g(X)], \mathbb{E}[\nu(X) \nu(X)']),       \\
	 & \sqrt{n}(\tilde{\theta}_n - \theta_0) \stackrel{g}{\rightsquigarrow} N(\mathbb{E}[\tau(X) g(X)], \mathbb{E}[\tau(X) \tau(X)']).
\end{align*}
Since $\nu \in \bar{T}(P)$, the asymptotic bias of $\hat{\theta}_n$ depends only on $\Pi_T(g)$, and thus the Hausman test cannot detect it. Moreover, when $g \in \bar{T}(P)$, both efficient and inefficient estimators share the same asymptotic bias. This occurs because $\psi(Q) = \varphi(Q)$ must hold for all $Q \in \mathbf{P}$.
That is, when the null hypothesis is true, the two estimators must estimate the same parameter.

We summarize the assumptions used to formalize the above conclusion and obtain the following proposition.

\begin{Assumption} \label{hausman_assumption}
	\begin{enumerate}
		\item $T(P)$ and $M(P)$ are linear.
		      
		\item The functionals $\psi$ and $\varphi$ are pathwise differentiable at $P$ with respect to $T(P)$ and $M(P)$, respectively.
		      
		\item $\hat{V}$ is consistent for $V$.
	\end{enumerate}
\end{Assumption}

\begin{Proposition} \label{hausman_properties}
	Suppose that $P$ is locally overidentified by $\mathbf{P}$ and that Assumption~\ref{hausman_assumption} holds. Let the Hausman test be based on \eqref{hausman_stat} with local asymptotic power function $\pi$ and asymptotic level $\alpha$.
	Then, $\hat{\theta}_n$ can be asymptotically biased only if $\Pi_T(g)\neq 0$, while $\tilde{\theta}_n$ can be asymptotically biased only if $\Pi_M(g)\neq 0$.
	Moreover, $\pi(g)>\alpha$ only if $\Pi_{T^\bot \cap M}(g)\neq 0$.
\end{Proposition}

Proposition~\ref{hausman_properties} reveals features of the Hausman test that are not immediately evident from the general results in Section~\ref{subsec:orthogonality}.
Namely, contrary to the conventional interpretation, the power of the Hausman test can be driven by the asymptotic bias of the inefficient estimator.
In the case where $g \in \bar{T}(P)^\bot \cap \bar{M}(P)$, the test may reject its null hypothesis precisely because the inefficient estimator is asymptotically biased, even though the efficient estimator remains asymptotically unbiased.

The result may seem incompatible with the standard setting of the Hausman test, which assumes that the inefficient estimator is valid under all scenarios.
The conventional interpretation becomes compatible with our result if one implicitly restricts attention to directions of local deviation under which the inefficient estimator remains asymptotically unbiased.
However, this effectively limits the analysis to a narrow range of deviations. 
We take an agnostic view of the directions of local deviation and analyze the behavior of estimators and test statistics over a broad class of deviations.

Another important feature of the Hausman test is that the estimators used to construct the test statistic determine the substantive null and alternative hypotheses.
The null and alternative hypotheses of the Hausman test are often left implicit in econometrics textbooks, giving readers a vague understanding of what the test formally evaluates.
Our analysis implies that the null model of the Hausman test is given by the model under which the estimator is asymptotically efficient.
For instance, in the case of the DWH test, where researchers compare the OLS estimator with the 2SLS estimator, the null model is the homoskedastic linear regression model, since OLS is asymptotically efficient under this model.
In contrast, the maintained model is the homoskedastic linear IV
model.
A detailed analysis of the DWH test is provided in Section~\ref{sec:dwh}.

It should be emphasized that, although the orthogonal decomposition of the score function follows directly from \cite{chen2018overidentification}, the implications developed in thie section are not a straightforward consequence of their analysis. In particular, \cite{chen2018overidentification} do not explicitly examine the asymptotic bias of estimators or its implications for estimator selection.


\section{$J$-test} \label{sec:jtest}
In this section, we examine the relationship between the efficient GMM estimator and the $J$-test within the framework introduced in Sections~\ref{sec:setup} and \ref{sec:main}. The purpose of this section is to clarify how our results relate to the existing literature. In particular, \cite{hall2005} provided an orthogonal decomposition of moment restrictions into identifying and overidentifying restrictions. We show that our decomposition of the score function corresponds exactly to his decomposition of moment restrictions.

To begin the analysis, we first reformulate the model following the approach of \cite{sueishi2024large}.
This formulation is convenient because it allows us to derive the tangent set in the conventional manner of the semiparametric estimation literature (e.g., Section 25.4 of \citealt{van1998asymptotic}).
Since model \eqref{iden_asm} involves two unknown parameters $\theta_0$ and $P$, we write the model as a set of distributions indexed by the finite-dimensional parameter $\theta \in \Theta$ and the infinite-dimensional nuisance parameter $\eta \in \mathcal{M}$.
Specifically, for given $\theta \in \Theta$ and $\eta \in \mathcal{M}$, we define $P_{\theta, \eta}$ as the solution to
\begin{eqnarray}\label{mc2}
	\min_{Q \in \mathbf{P}_\theta}  \int   \log \frac{dQ}{d\eta} dQ,
\end{eqnarray}
where $\mathbf{P}_\theta =\{ Q \in \mathcal{M} : \int m_\theta dQ =0\}$.
That is, $P_{\theta, \eta}$ is the projection of $\eta$ onto the set $\mathbf{P}_\theta$ with respect to the Kullback--Leibler divergence.
By a duality theorem, $P_{\theta, \eta}$ satisfies
\[
	\frac{dP_{\theta, \eta}}{d\eta} = \frac{\exp(\lambda_{\theta, \eta}' m_\theta)}{\int \exp(\lambda_{\theta, \eta}' m_\theta) d\eta},
\]
where $\lambda_{\theta, \eta} = \arg \min_{\lambda \in \mathbb{R}^l} \int \exp(\lambda'm_\theta) d\eta$.
See \cite{borwein1991duality} and \cite{komunjer2016existence} for details.
Our model $\mathbf{P}=\{P_{\theta, \eta} :\theta \in \Theta, \eta \in \mathcal{M} \}$ is the set of all distributions on $\mathcal{X}$ that satisfy moment restrictions for some $\theta \in \Theta$.

We now specify the tangent set of $\mathbf{P}$ at $P$. 
To do this, we consider a path of the form $P_t = P_{\theta_0+t h, \eta_t}$, where $h \in \mathbb{R}^p$ and $\eta_t$ is a perturbation from $P$ that coincides with $P$ at $t=0$.
Under certain conditions, $P_t$ satisfies
\[
	\lim_{t \to 0} \int \left( \frac{dP_t^{1/2} -dP^{1/2}}{t} -\frac{1}{2} (h'\dot{\ell}_{\theta_0, \eta_0} + \dot{l} ) dP^{1/2} \right)^2 =0
\]
where
\[
	\dot{\ell}_{\theta_0, \eta_0}(x)= - \mathbb{E}[\nabla m_{\theta_0}(X) ]' \Sigma^{-1} m_{\theta_0}(x)
\]
with $\nabla m_\theta =\partial m_\theta/ \partial \theta' $ and $\Sigma = \mathbb{E}[m_{\theta_0}(X)m_{\theta_0}(X)']$. 
Moreover, $\dot{l}: \mathcal{X} \to \mathbb{R}$ is an element of the set $\dot{\mathbf{P}}_\eta = \{ \dot{l} \in L_0^2(P) : \mathbb{E}[m_{\theta_0}(X) \dot{l}(X)] =0 \}$.
See \cite{sueishi2024large} for details.
The function $\dot{\ell}_{\theta_0, \eta_0}$ is interpreted as the score function for $\theta_0$ when $\eta_0$ is fixed, while $\dot{l}$ is interpreted as the score function for $\eta_0$ when $\theta_0$ is fixed.
The tangent set is given by $T(P) = \{ {\rm lin} \ \dot{\ell}_{\theta_0, \eta_0} + \dot{\mathbf{P}}_\eta \}$, where ${\rm lin}$ denotes the linear span.

Notice that $\dot{\ell}_{\theta_0, \eta_0}$ is orthogonal to all elements of $\dot{\mathbf{P}}_\eta$.
This orthogonality confirms that $\dot{\ell}_{\theta_0, \eta_0}$ is indeed the efficient score function for estimating $\theta_0$.
Consequently, the efficient information matrix is given by
\[
	I_{\theta_0, \eta_0}
	= \mathbb{E}[\dot{\ell}_{\theta_0, \eta_0}(X) \dot{\ell}_{\theta_0, \eta_0}(X)']
	= \mathbb{E}\left[ \nabla m_{\theta_0}(X) \right]'
	\Sigma^{-1} \mathbb{E}\left[ \nabla m_{\theta_0}(X) \right].
\]
The efficient influence function is therefore $I_{\theta_0, \eta_0}^{-1} \dot{\ell}_{\theta_0, \eta_0}$.

Now, we investigate the local asymptotic property of the GMM estimator.
The efficient GMM estimator satisfies
\[
	\sqrt{n}(\hat{\theta}_n - \theta_0) = \frac{1}{\sqrt{n}} \sum_{i=1}^n I_{\theta_0, \eta_0}^{-1} \dot{\ell}_{\theta_0, \eta_0}(X_i) + o_P(1).
\]
This demonstrates that the efficient GMM estimator is best regular in our model $\mathbf{P}$.
Moreover, it follows from Le Cam's third lemma that
\begin{equation}\label{gmm_local}
	\sqrt{n}(\hat{\theta}_n - \theta_0) \stackrel{g}{\rightsquigarrow} N( I_{\theta_0, \eta_0}^{-1} \mathbb{E}[\dot{\ell}_{\theta_0, \eta_0}(X) g(X)], I_{\theta_0, \eta_0}^{-1})
\end{equation}
for any $g \in L_0^2(P)$. 
Here, the efficient influence function clearly belongs to $\bar{T}(P)$ and is therefore orthogonal to $\Pi_{T^\bot}(g)$.
Thus, the asymptotic bias depends only on $\Pi_T(g)$.

The GMM estimator can be asymptotically unbiased even when moment restrictions are violated because it utilizes only a part of the moment restrictions.
It follows from \eqref{gmm_local} that the GMM estimator is asymptotically unbiased if
\begin{equation}\label{gmm_bias}
	\mathbb{E}[\nabla m_{\theta_0}(X)]' \Sigma^{-1} \mathbb{E}[m_{\theta_0}(X) g(X)] =0.
\end{equation}
Although $\mathbb{E}[m_{\theta_0}(X) g(X)] \neq 0$ in general, the left-hand side of \eqref{gmm_bias} can be zero because the rank of $\mathbb{E}[\nabla m_{\theta_0}(X)]' \Sigma^{-1}$ is $p$.
Note that the left-hand side of \eqref{gmm_bias} is approximately the derivative of the GMM population objective function under local deviation $P_{1/\sqrt{n},g}$. Evidently, if the objective function is minimized at $\theta_0$, then the GMM estimator is valid even if the moment restrictions are not satisfied at $\theta_0$.

Next, we investigate the local asymptotic properties of the $J$-test statistic.
It can be expressed as
\[
	J_n = \left( \frac{1}{\sqrt{n}} \sum_{i=1}^n m_{\hat{\theta}_n}(X_i) \right)' \Sigma^{-1}
	\left( \frac{1}{\sqrt{n}} \sum_{i=1}^n m_{\hat{\theta}_n}(X_i) \right) +o_P(1).
\]
Some calculation yields that
\[
	\Sigma^{-1/2}\frac{1}{\sqrt{n}} \sum_{i=1}^n m_{\hat{\theta}_n}(X_i)
	\stackrel{g}{\rightsquigarrow} N((I-P(\theta_0)) \Sigma^{-1/2} \mathbb{E}[m_{\theta_0}(X)g(X)], I-P(\theta_0)),
\]
where $P(\theta_0) =\Sigma^{-1/2} \mathbb{E}[\nabla m_{\theta_0}(X)]  \left( \mathbb{E}[\nabla m_{\theta_0}(X)]' \Sigma^{-1}  \mathbb{E}[\nabla m_{\theta_0}(X)]  \right)^{-1}
	\mathbb{E}[\nabla m_{\theta_0}(X)]' \Sigma^{-1/2}$ is a projection matrix.
It follows that $J_n$ converges in distribution to the noncentral chi-square distribution with degrees of freedom $l-p$ and the noncentrality parameter
\[
	\mathbb{E}[m_{\theta_0}(X) g(X) ]' \Sigma^{-1/2} (I-P(\theta_0)) \Sigma^{-1/2} \mathbb{E}[m_{\theta_0}(X) g(X)].
\]
If $g \in \bar{T}(P)$, then $g$ can be written as $g = h'\dot{\ell}_{\theta_0, \eta_0} + \dot{l}$ for some $h \in \mathbb
	{R}^p$ and $\dot{l} \in \dot{\mathbf{P}}_\eta$. In this case, we have
\begin{equation}\label{proj}
	(I-P(\theta_0))\Sigma^{-1/2} \mathbb{E}[m_{\theta_0}(X)g(X)]
	= (I-P(\theta_0)) \Sigma^{-1/2}  \mathbb{E}[\nabla m_{\theta_0}(X)]h = 0,
\end{equation}
which implies that the noncentrality parameter is zero. 
Therefore, the $J$-test is locally unbiased for testing $H_0: \mu \in \mathbf{P}$ against $H_1: \mu \in \mathcal{M} \setminus \mathbf{P}$.
In particular, it fails to detect the asymptotic bias in the efficient GMM estimator.

\cite{hall2005} showed a similar orthogonality result from a different perspective.  He considered a local deviation of the form:
\[
	\Sigma^{-1/2} \mathbb{E}_n[m_{\theta_0}(X)] = \frac{\delta}{\sqrt{n}}
\]
for some $\delta \in \mathbb{R}^l$, where $\mathbb{E}_n$ is the expectation with respect to the local deviation.
The vector $\Sigma^{-1/2} \mathbb{E}_n[m_{\theta_0}(X)]$ can be decomposed as
\begin{equation}\label{decomp_hall}
	P(\theta_0) \Sigma^{-1/2} \mathbb{E}_n[m_{\theta_0}(X)]
	+(I-P(\theta_0))  \Sigma^{-1/2} \mathbb{E}_n[m_{\theta_0}(X)].
\end{equation}
Since $P(\theta_0)$ is a projection matrix, the two terms in \eqref{decomp_hall} are orthogonal to each other.
\cite{hall2005} defines that the identifying restrictions are satisfied if the first term of \eqref{decomp_hall} is zero and the overidentifying restrictions are satisfied if the second term is zero.
He showed that the GMM estimator is asymptotically biased only if the identifying restrictions are locally violated, while the $J$-test has nontrivial local power only if the overidentifying restrictions are locally violated.

Our decomposition of the score function essentially yields the same result as above. If the expectation is taken with respect to the local deviation $P_{1/\sqrt{n},g}$, then $\mathbb{E}_n[m_{\theta_0}(X)]$ is asymptotically equivalent to $\mathbb{E}[m_{\theta_0}(X)g(X)]/\sqrt{n}$. The correspondence between the two frameworks is thus direct: The first term of (\ref{decomp_hall}) can be nonzero only if $\Pi_T(g) \neq 0$, since \eqref{gmm_bias} holds if $g \in \bar{T}(P)^\bot$. Conversely, the second term of (\ref{decomp_hall}) can be nonzero only if $\Pi_{T^\bot}(g) \neq 0$, as \eqref{proj} holds if $g \in \bar{T}(P)$.
Therefore, the decomposition of the score function produces the same orthogonality result as \cite{hall2005}.

\begin{Remark}
	\cite{newey1985generalized} showed that the $J$-test lacks power against certain local deviations.
	For the weighting matrix $W=\Sigma^{-1}$, the condition in Proposition 1 of \cite{newey1985generalized} is written as
	\[
		(I-P(\theta_0)) \Sigma^{-1/2} \mathbb{E}[m_{\theta_0}(X)g(X)] =0,
	\]
	which corresponds exactly to the case $g \in \bar{T}(P)$ in our framework.
	Our orthogonal decomposition clarifies why his condition implies lack of power.
	While \cite{newey1985generalized} considers only finite-dimensional deviation from $P$, the fundamental nature of the analysis is the same as ours.
\end{Remark}

\begin{Remark}\label{guggenberger}
	Our orthogonality result has direct implications for two-stage testing procedures. \cite{guggenberger2012size} showed that in the homoskedastic linear IV model, the asymptotic size of a two-stage test---where the Sargan test is conducted first, followed by a 2SLS-based $t$-test if the Sargan test does not reject---can be as large as $1-\varepsilon_P$, where $\varepsilon_P$ is the nominal size of the Sargan test.
	
	Building on our analysis, we can extend their result to more general models.
	We give a sketch of the argument.
	Let $\theta$ be a scalar parameter and consider testing $H_0: \theta = \theta_0$ against $H_1: \theta \neq \theta_0$.
	Suppose that the DGP corresponds to $\theta_0$, so that the null hypothesis is true, but that the moment restrictions are locally violated with 
	score $g = h\dot{\ell}_{\theta_0,\eta_0} + \dot{l}$ for $h \in \mathbb{R}$ and $\dot{l} \in \dot{\mathbf{P}}_\eta$.
	Since the $J$-statistic converges to a central chi-square distribution, the $J$-test fails to reject its null with probability approaching $1-\varepsilon_P$.
	On the other hand, the efficient GMM estimator satisfies $\sqrt{n}(\hat{\theta}_n - \theta_0) \stackrel{g}{\rightsquigarrow} N(h,1/I_{\theta_0, \eta_0})$.
	By choosing $h$ sufficiently large, the probability that the second-stage $t$-test rejects the null hypothesis can be made arbitrarily close to one. Hence, the asymptotic size of the two-stage test is bounded below by $1-\varepsilon_P$.

\end{Remark}

\begin{Remark}
	As noted in \cite{andrews2025purpose}, empirical researchers often use inefficient GMM estimators whose weighting matrices are chosen in an arbitrary manner.  
	For such estimators, the orthogonality result does not hold, since they may exhibit bias even when $g \in T(P)^\bot$.  
	Although the $J$-test cannot accurately detect bias for any specific estimator, \cite{andrews2025purpose} propose a method that employs the $J$-statistic to determine the set of point estimates obtainable by varying the weighting matrix, conditional on a given standard error.
	Therefore, while the $J$-statistic may not be informative for pretesting, reporting it can still be useful for other purposes.

\end{Remark}


\section{DWH test} \label{sec:dwh}

This section investigates the properties of the DWH test.
The test statistic is given by
\[
	T_n =n(\tilde{\theta}_n-\hat{\theta}_n )' \hat{V}^{-} (\tilde{\theta}_n-\hat{\theta}_n),
\]
where $\hat{\theta}_n$ and $\tilde{\theta}_n$ denote the OLS and 2SLS estimators for $\theta_0 \in \mathbb{R}^p$ in the linear model:
\[
	Y =X'\theta_0  + e = X_1'\theta_{01} + X_2'\theta_{02} +e,
\]
where $X_1$ is suspected to be endogenous.
Here, $\hat{V}^{-}$ denote the generalized inverse of a consistent estimator for the asymptotic variance of $\tilde{\theta}_n-\hat{\theta}_n$.

As stated in Section~\ref{subsec:hausman}, the estimators used to construct the test statistic define the substantive null and alternative hypotheses.  
Let $Z_1$ be a vector of instrumental variables for $X_1$ and define $Z=(Z_1', X_2')'$.  
The null and maintained models are specified as the sets of joint distributions of $(X_1', Y, Z')'$.
Since the OLS estimator is efficient in the homoskedastic linear regression model and the 2SLS estimator is efficient in the homoskedastic linear IV model, we consider the following models:
\begin{align*}
	\mathbf{P} & = \left\{Q \in \mathcal{M}: \mathbb{E}_Q[Y-X'\theta \mid X_1, Z]=0 \ \text{and} \ \mathbb{E}_Q[(Y-X'\theta)^2 \mid X_1, Z]=\sigma^2 \ \text{for some} \ \theta \ \text{and} \ \sigma^2 \right\}, \\
	\mathbf{M} & = \left\{Q \in \mathcal{M}: \mathbb{E}_Q[Z(Y-X'\theta)]=0 \ \text{and} \ \mathbb{E}_Q[(Y-X'\theta)^2 \mid Z]=\sigma^2 \ \text{for some} \ \theta \ \text{and} \ \sigma^2 \right\},
\end{align*}
where $\mathbb{E}_Q[\cdot]$ and $\mathbb{E}_Q[\cdot|\cdot]$ denote the unconditional and conditional expectations with respect to $Q \in \mathcal{M}$.
The parameter of interest $\theta_0$ satisfies
\[
	\mathbb{E}[Y-X'\theta_0 \mid X_1,Z] = 0.
\]

It is important to note that the set of conditioning variables in $\mathbf{P}$ is $(X_1', Z')'$ rather than $X$.
While conditioning only on $X$ is sufficient if we do not consider the maintained model, 
including $Z_1$ is necessary when the maintained model exists to ensure that the exclusion restriction is satisfied in both models.
Without the exclusion restrictions, the two estimators may estimate different parameters even under the null hypothesis.

The tangent set of $\mathbf{M}$ can be obtained by using the result of Section~\ref{sec:jtest} with $m_{\theta}(x_1,y,z) =z(y-x'\theta)$.
If we ignore the homoskedasticity assumption, the efficient score for estimating $\theta_0$ is given by $\dot{\ell}_{\theta_0, \eta_0}^{M}(x_1, y,z)=\mathbb{E}[XZ']\mathbb{E}[ZZ'e^2]^{-1}z(y-x'\theta_0)$.
Under homoskedasticity, we have $\mathbb{E}[ZZ'e^2]=\sigma_0^2 \mathbb{E}[ZZ']$ for some constant $\sigma_0^2$.
Hence, the efficient score function reduces to 
\[
	\dot{\ell}_{\theta_0, \eta_0}^M(x_1, y, z) = \mathbb{E}[XZ'] \mathbb{E}[ZZ']^{-1} z(y-x'\theta_0)/\sigma_0^2.
\]
The tangent set is given by $M(P)=\{ {\rm lin} \ \dot{\ell}_{\theta_0,\eta_0}^M + \dot{\mathbf{M}}_\eta \}$, where 
\[
	\dot{\mathbf{M}}_\eta =\left\{ \dot{l}^M \in L_0^2(P): \mathbb{E}[Ze \dot{l}^M(X_1,Y,Z)] =0  \right\}.
\]

To obtain the tangent set of $\mathbf{P}$, we introduce a new formulation of the conditional moment restriction models.
As in the case of $\mathbf{M}$, we first rewrite the model as the set of distributions indexed by the parameter of interest $\theta \in \Theta$ and an infinite-dimensional nuisance parameter $\eta \in \mathcal{M}$.
Then, our model can be written in the form $\mathbf{P}=\{P_{\theta, \eta} : \theta \in \Theta, \eta \in \mathcal{M} \}$.
Using an argument similar to that in Section~\ref{sec:jtest}, we obtain the efficient score function: $\dot{\ell}_{\theta_0, \eta_0}^P(x_1, y, z) = x(y-x'\theta_0)/\mathbb{E}[e^2|X_1=x_1, Z=z]$.
See the Appendix for the derivation.
Under homoskedasticity, this reduces to 
\[
	\dot{\ell}_{\theta_0,\eta_0}^P(x_1, y,z) = x (y-x'\theta_0)/\sigma_0^2,
\]
which confirms that the OLS estimator is best regular in model $\mathbf{P}$.
The tangent set is given by $T(P)=\{ {\rm lin} \ \dot{\ell}_{\theta_0, \eta_0}^P + \dot{\mathbf{P}}_\eta\}$, where
\[
	\dot{\mathbf{P}}_\eta = \left\{ \dot{l}^P \in L_0^2(P) : \mathbb{E}[a(X_1, Z) e \dot{l}^P(X_1,Z, Y)]=0 \ \text{for any function} \ a \ \text{of} \ (X_1, Z) \right\}.
\]

We now show that the DWH test is locally unbiased for testing 
\begin{equation*}
	H_0: \mu \in \mathbf{P} \quad \text{vs.} \quad H_1: \mu \in \mathbf{M} \setminus \mathbf{P}.
\end{equation*}
By Le Cam's third lemma, the asymptotic distributions of the two estimators under local deviations are:
\begin{align*}
	\sqrt{n}(\hat{\theta}_n-\theta_0)
	 & \stackrel{g}{\rightsquigarrow}
	N\!\left(
	\mathbb{E}[XX']^{-1}
	\mathbb{E}[Xe\, g(X_1,Y,Z)],
	\;
	\sigma_0^2 \mathbb{E}[XX']^{-1}
	\right),                          \\
	\sqrt{n}(\tilde{\theta}_n-\theta_0)
	 & \stackrel{g}{\rightsquigarrow}
	N\!\Bigg(
	(\mathbb{E}[XZ']\mathbb{E}[ZZ']^{-1}\mathbb{E}[ZX'])^{-1}
	\mathbb{E}[XZ']\mathbb{E}[ZZ']^{-1}
	\mathbb{E}[Ze\, g(X_1,Y,Z)],
	\\
	 & \qquad \qquad
	\sigma_0^2
	(\mathbb{E}[XZ']\mathbb{E}[ZZ']^{-1}\mathbb{E}[ZX'])^{-1}
	\Bigg).
\end{align*}
We consider three cases.

\begin{description}
	\item[Case 1: $g \in \bar{T}(P)$.] 
	      Suppose first that $g \in \bar{T}(P)$. Then $g$ can be written as $g = h' \dot{\ell}_{\theta_0, \eta_0}^P + \dot{l}^P$ for some $h \in \mathbb{R}^p$ and $\dot{l}^P \in \dot{\mathbf{P}}_\eta$.
	      The asymptotic bias term for the OLS estimator simplifies to
	      \[
		      \mathbb{E}[XX']^{-1} \mathbb{E}[Xe g(X_1, Y, Z)] =
		      \mathbb{E}[XX']^{-1} \mathbb{E}[XX' e^2] h / \sigma_0^2 = h.
	      \]
	      Similarly, the asymptotic bias term for the 2SLS estimator simplifies to $h$.
	      Thus, the OLS and 2SLS estimators have the same asymptotic bias, and consequently, the DWH test has no local power.
	      
	      This result is intuitive: if $\mathbb{E}_\mu[Y - X'\theta \mid X_1, Z] = 0$ holds for some $\theta$, then $\mathbb{E}_\mu[Z(Y - X'\theta)] = 0$ must also hold for the same \(\theta\). It would therefore be incoherent to regard 2SLS as asymptotically unbiased while OLS is biased; they must share the same asymptotic bias when $g \in \bar{T}(P)$.
	      
	\item[Case 2: $g \in \bar{T}(P)^\perp \cap \bar{M}(P)$.] 
	      Next, suppose that \(g \in \bar{T}(P)^\perp \cap \bar{M}(P)\). 
	      In this case, $g$ is orthogonal to \(\dot{\ell}_{\theta_0, \eta_0}^P\), and thus the OLS estimator is asymptotically unbiased.
	      At the same time, \(g\) can be expressed as $g = h' \dot{\ell}_{\theta_0, \eta_0}^M + \dot{l}^M$ for some \(h \in \mathbb{R}^p\) and \(\dot{l}^M \in \dot{\mathbf{M}}_\eta\), 
	      which implies that the 2SLS estimator may be asymptotically biased with bias $h$.
	      Since the two estimators have different asymptotic means when $h \neq 0$, the DWH test has nontrivial local power. It is important to note that rejection of the null hypothesis occurs due to the asymptotic bias of the 2SLS estimator, even when the OLS estimator is asymptotically unbiased.
	      
	\item[Case 3: $g \in \bar{M}(P)^\perp$.] 
	      Finally, suppose that \(g \in \bar{M}(P)^\perp\). Because \(g\) is orthogonal to both 
	      \(\dot{\ell}_{\theta_0,\eta_0}^P\) and \(\dot{\ell}_{\theta_0, \eta_0}^M\), it does not 
	      contribute to the asymptotic bias of either estimator. Therefore, the DWH test does not 
	      have local power.
\end{description}

Three cases above should not be interpreted as mutually exclusive scenarios. A general local deviation $g \in L_0^2(P)$ need not belong entirely to any one of these subspaces. Rather, $g$ can be decomposed into orthogonal components associated with $\bar{T}(P)$, $\bar{T}^\bot \cap \bar{M}(P)$, and $\bar{M}(P)^\bot$.
The preceding discussion examines each component in isolation, thereby revealing the distinct mechanisms through which local deviation influences the asymptotic behavior of the OLS and 2SLS estimators and the resulting local power of the DWH test.

Finally, we explain how the OLS estimator can be asymptotically unbiased when $g \in \bar{T}(P)^{\bot}$.
Because the null model is misspecified in this case, there is no $\theta$ that satisfies
$\mathbb{E}_\mu[Y-X'\theta|X_1, Z]=0$ and $\mathbb{E}_\mu[(Y-X'\theta)^2|X_1, Z]=\sigma^2$.
However, the misspecification in the conditional moments does not necessarily exclude the key condition required for the validity of OLS: $\mathbb{E}_\mu[X(Y-X'\theta_0)]=0$. 
Thus, OLS can be asymptotically unbiased.
In this case, $X_1$ is uncorrelated with $e$ under $\mu$ even though the null hypothesis is not true.
Therefore, the DWH test does not necessarily serve as a test of ``exogeneity.”


\section{Simulations}\label{sec:simulation}

This section presents Monte Carlo results for the $J$-test and the DWH test. The theoretical results above are based on local misspecification, but the same logic can be illustrated in simple linear designs using fixed misspecification, where the distance between the DGP and the benchmark distribution does not shrink as the sample size increases. In this setting, asymptotic bias appears as inconsistency, and power appears as rejection with high probability. The simulations show that a specification test may reject even when the associated estimator is consistent, and may fail to reject even when it is inconsistent.

\subsection{$J$-test}
This subsection investigates the relationship between the $J$ (Sargan) test and the 2SLS estimator.
We draw a random sample $\{ Y_i, X_i, Z_{1i}, Z_{2i}\}_{i=1}^n
$ generated according to the following system:
\begin{align*}
	Y_i & = \theta_0 X_i + u_i,                       \\
	X_i & = \alpha_1 Z_{1i} + \alpha_2 Z_{2i} + v_i,  \\
	u_i & = \delta_1 Z_{1i} + \delta_2 Z_{2i} + e_i.
\end{align*}
The potential instruments $Z_{1i}$ and $ Z_{2i}$ are mutually independent and follow standard normal distributions.
The error terms \((v_i, e_i)'\) are generated from a bivariate normal distribution with mean zero, unit variances, and correlation $0.5$.
We consider three DGPs (DGP1–DGP3) corresponding to the settings below.
\begin{itemize}
	\item $\alpha_1=1$, $\alpha_2=1$, $\delta_1=0$, $\delta_2=0 \;\; (\text{DGP1})$
	\item $\alpha_1=1$, $\alpha_2=0$, $\delta_1=0$, $\delta_2=0.2 \;\; (\text{DGP2})$
	\item $\alpha_1=1$, $\alpha_2=1$, $\delta_1=0.2$, $\delta_2=0.2 \;\; (\text{DGP3})$
\end{itemize}

DGP1 corresponds to the ideal distribution $P$, while DGP2 and DGP3 represent deviations from $P$.
For DGP2, there is no value of $\theta$ that makes the instruments orthogonal to $u_i$.
Nevertheless, the population objective function of the 2SLS estimator attains its minimum at $\theta_0$, the true parameter value used to generate the data.
Consequently, we expect that the 2SLS estimator is valid, whereas the $J$-test rejects the null hypothesis with high probability.
In contrast, DGP3 is designed so that the moment restrictions are not satisfied at $\theta_0$ but are satisfied at $\theta = \theta_0 + 0.2$.
Accordingly, we expect that the $J$-test does not reject the null hypothesis, whereas the 2SLS estimator is invalid.
This corresponds to the case discussed in Remark~\ref{guggenberger}.

Table~\ref{tab:sargan_results} reports the simulation results based on 1,000 iterations with a sample size of $n = 500$.
The true parameter is set to $\theta_0 = 1.5$, and the $J$-test is implemented at the $5\%$ nominal size.
We report the mean of the 2SLS estimator, the coverage rate of the associated $95\%$ confidence interval, and the rejection rate of the $J$-test.

\begin{table}[htbp]
	\centering
	\caption{$J$-test and 2SLS estimator}
	\vspace{6pt}
	\label{tab:sargan_results}
	\begin{tabular}{lccc}
		\hline\hline
		     & Mean  & Coverage Rate & Rejection Rate \\
		\hline
		DGP1 & 1.501 & 0.943         & 0.061          \\
		DGP2 & 1.500 & 0.944         & 0.991          \\
		DGP3 & 1.701 & 0.000         & 0.057          \\
		\hline
	\end{tabular}
\end{table}

The results highlight the contrast between the 2SLS estimator and the $J$-test across DGP2 and DGP3: the 2SLS estimator remains valid under DGP2 but is invalid under DGP3, whereas the 
$J$-test detects the deviation under DGP2 but fails to reject under DGP3, consistent with our theoretical expectations.

\subsection{DWH test}
This subsection investigates the relationship between the DWH test and the OLS estimator.
We draw a random sample $\{ Y_i, X_{1i}, X_{2i}, Z_i \}_{i=1}^n$ generated according to the following system:
\begin{align*}
	Y_i    & = \theta_{10} X_{1i} + \theta_{20} X_{2i} + u_i, \\
	X_{1i} & = 0.8 Z_i + v_i,                                 \\
	u_i    & = \delta Z_i + e_i,
\end{align*}
where $X_{2i}$ and $Z_i$ are mutually independent standard normal random variables. The error terms $(v_i, e_i)'$ are generated from a bivariate normal distribution with mean zero, unit variances, and correlation $\rho_{ve}$. The variable $Z_i$ serves as a potential instrument for $X_{1i}$. 
We consider three DGPs corresponding to the settings below.
\begin{itemize}
	\item $\delta=0, \rho_{ve}=0 \;\; (\text{DGP1})$
	\item $\delta=0.2, \rho_{ve}=-0.16 \;\; (\text{DGP2})$
	\item $\delta=0.2, \rho_{ve}=0.25 \;\; (\text{DGP3})$
\end{itemize}

DGP1 corresponds to the ideal distribution $P$, while DGP2 and DGP3 represent deviations from $P$.
For DGP2, $\rho_{ve}$ is chosen so that the covariance between $X_{1i}$ and $u_i$ is zero, whereas the instrument $Z_i$ is not valid.
Consequently, we expect the DWH test to reject the null hypothesis due to the invalidity of the 2SLS estimator, even though the OLS estimator remains valid.
For DGP3, both $X_{1i}$ and $Z_i$ are correlated with $u_i$.
We set $\rho_{ve}$ so that the OLS and 2SLS estimators share the same probability limit.
Therefore, we expect the DWH test not to reject the null hypothesis, despite the fact that both the OLS and 2SLS estimators are invalid.

Table~\ref{tab:dwh_results} reports the simulation results based on 1,000 iterations with a sample size of $n=500$.
In this simulation, the DWH test is implemented using the equivalent augmented-regression form for numerical stability, in which the residual from the first-stage regression of $X_{1i}$ on $(Z_i,X_{2i})'$ is included in the structural equation and its coefficient is tested.
We set $\theta_{10} =\theta_{20}= 1.5$, and the DWH test is conducted at the 5\% nominal size.
We report the mean of the OLS estimator for $\theta_{10}$, the coverage rate of the associated $95\%$ confidence interval, and the rejection rate of the DWH test. 
The results are again consistent with our theoretical predictions.

\begin{table}[htbp]
	\centering
	\caption{DWH test and OLS estimator}
	\vspace{6pt}
	\label{tab:dwh_results}
	\begin{tabular}{lccc}
		\hline\hline
		     & Mean  & Coverage Rate & Rejection Rate \\
		\hline
		DGP1 & 1.501 & 0.957         & 0.038          \\
		DGP2 & 1.500 & 0.950         & 1.000          \\
		DGP3 & 1.750 & 0.000         & 0.054          \\
		\hline
	\end{tabular}
\end{table}

The simulation settings described above are deliberately stylized and are chosen to make the results transparent.
However, it is entirely plausible to encounter cases in practice where an estimator exhibits only a small bias yet the null hypothesis is rejected, or where an estimator is invalid but the test cannot reject the null hypothesis.

\section{Conclusion}\label{sec:conclusion}

This paper studied the local asymptotic properties of specification tests and asymptotically efficient estimators in a broad class of semiparametric models. We showed that the directions of local deviation detectable by locally unbiased specification tests are orthogonal to those that induce asymptotic bias in asymptotically efficient estimators. As a result, specification tests alone cannot generally be used to assess the asymptotic validity of efficient estimators. Furthermore, Hausman tests not only provide limited information about the bias of efficient estimators, but may also lead researchers to select inefficient estimators that suffer from asymptotic bias.

Our results should not be interpreted as suggesting that specification tests are completely useless. Rather, our results clarify the precise sense in which specification tests are informative and the limitations of the conclusions that can be drawn from them. In particular, interpreting the outcome of a specification test as evidence for or against the validity of an estimator requires additional assumptions that are often left implicit in empirical applications. By making these assumptions explicit, our analysis clarifies the conditions under which specification tests can be used to draw conclusions about estimator validity.

\appendix
\section{Appendix}

The Appendix derives the tangent set for conditional moment restriction models.
The derivation closely follows that of the unconditional moment restriction models considered by \cite{sueishi2024large}.

Let $P$ be the joint probability distribution of $(X, W)$ whose support is $\mathcal{X} \times \mathcal{W}$, and let $\mathcal{M}$ be the set of all distributions on $\mathcal{X} \times \mathcal{W}$.
We write $P =P_{W|X} P_X$, where $P_{W|X}$ is the conditional distribution of $W$ given $X$, and $P_X$ is the marginal distribution of $X$. 
The parameter of interest $\theta_0 \in \Theta \subset \mathbb{R}^p$ is a unique vector that satisfies
\begin{equation}\label{conditional_moment}
	0 = \mathbb{E}[m_{\theta_0}(W)|X] = \int m_{\theta_0} dP_{W|X} \ \text{a.s.} \ P_X,
\end{equation}
where $m : \mathcal{W} \times \Theta \to \mathbb{R}^l$ is a known vector-valued function.

In practice, $X$ and $W$ share common elements, so the integral in \eqref{conditional_moment} is effectively taken with respect to the components of $W$ that are not included in $X$. For example, in the linear regression model $Y = X'\theta_0 + e$, we have $W = (Y, X')'$ and $m_\theta(W) = Y - X'\theta$, so that
\[
	\mathbb{E}[Y - X'\theta_0| X] = \int (y - X'\theta_0)  dP_{Y\mid X}(y \mid X).
\]
Although the notation in \eqref{conditional_moment} is somewhat imprecise, we adopt it for the sake of notational simplicity.

We present the model as a set of distributions on $\mathcal{X} \times \mathcal{W}$ indexed by the finite-dimensional parameter $\theta \in \Theta$ and the infinite-dimensional nuisance parameter $\eta \in \mathcal{M}$.
Write $\eta=\eta_{W|X}\eta_X$.
For each fixed $x$, we define $P_{\theta,\eta_{W|X}}(\cdot|x)$ as the conditional distribution that is closest to $\eta_{W|X}(\cdot|x)$ in Kullback--Leibler divergence among conditional distributions satisfying the moment restriction at $x$.
That is, $P_{\theta,\eta_{W|X}}(\cdot|x)$ solves
\[
	\min_{Q_{W|X}(\cdot|x)}
	\int \log \frac{dQ_{W|X}(w|x)}{d\eta_{W|X}(w|x)} dQ_{W|X}(w|x)
\]
subject to $\int m_\theta(w) dQ_{W|X}(w|x) =0$.
We then combine these conditional distributions with the marginal distribution $\eta_X$ and define
\[
	P_{\theta,\eta}=P_{\theta,\eta_{W|X}}\eta_X.
\]
The model is written as $\mathbf{P}=\{P_{\theta, \eta}: \theta \in \Theta, \eta \in \mathcal{M} \,\}$.

By a duality theorem, the conditional distribution $P_{\theta,\eta_{W|X}}$ satisfies
\[
	\frac{dP_{\theta, \eta_{W|X}}}{d\eta_{W|X}}(w|x) = \frac{\exp(\lambda_{\theta ,\eta_{W|X}}(x)'m_\theta(w))}{\int \exp(\lambda_{\theta, \eta_{W|X}}(x)'m_\theta(w)) d \eta_{W|X}(w|x)},
\]
where
\[
	\lambda_{\theta, \eta_{W|X}}(x) = \arg \min_{\lambda \in \mathbb{R}^l} \int \exp(\lambda' m_\theta(w)) d\eta_{W|X}(w|x).
\]
See, for instance, \cite{komunjer2016existence} for a rigorous argument.
Since $P_{\theta,\eta}$ and $\eta$ have the same marginal distribution of $X$, we have
\[
	\frac{dP_{\theta, \eta}}{d\eta}(x,w) = \frac{\exp(\lambda_{\theta, \eta_{W|X}}(x)'m_\theta(w))}{\int \exp(\lambda_{\theta,\eta_{W|X}}(x)'m_\theta(w)) d\eta_{W|X}(w|x)}.
\]
Let $\lambda_\theta(x) =\lambda_{\theta, P_{W|X}}(x)$.
Then, $\lambda_{\theta_0}(X)=0$ a.s. $P_X$.
Hence,
\begin{eqnarray*}
	\left. \frac{\partial}{\partial \theta} \log dP_{\theta, \eta_0}(x,w) \right|_{\theta=\theta_0} =
	\left[ \left. \frac{\partial \lambda_{\theta}(x)}{\partial \theta'} \right|_{\theta=\theta_0} \right]' m_{\theta_0}(w).
\end{eqnarray*}
Moreover, by the implicit function theorem,
\[
	\left. \frac{\partial \lambda_{\theta}(x)}{\partial \theta'} \right|_{\theta=\theta_0} = -\mathbb{E} \left[ m_{\theta_0}(W) m_{\theta_0}(W)' |X=x\right]^{-1} \mathbb{E} [\nabla m_{\theta_0}(W)|X=x].
\]
Thus,
\begin{align*}
	\dot{\ell}_{\theta_0, \eta_0}(x,w)
	 & = \left. \frac{\partial}{\partial \theta} \log dP_{\theta, \eta_0}(x,w) \right|_{\theta=\theta_0} \\
	 & = -\mathbb{E}[\nabla m_{\theta_0}(W)|X=x]' 
	\mathbb{E}[m_{\theta_0}(W)m_{\theta_0}(W)'|X=x]^{-1} m_{\theta_0}(w).
\end{align*}

We now derive the nuisance tangent set.
Consider a path of the form
\[
	P_t = P_{\theta_0 + ht, \eta_t}
\]
for $t \in [0, \epsilon)$, where $h \in \mathbb{R}^p$ and $\eta_t$ is a perturbation of $\eta_0=P$.
Because $P_{\theta_0,\eta_t}$ satisfies the conditional moment restriction, we have
\[
	\int \int m_{\theta_0}(w) a(x) dP_{\theta_0, \eta_t}(x, w) =0
\]
for any $t$ and for any function $a:\mathcal{X} \to \mathbb{R}$.
Differentiating this equality at $t=0$ gives
\[
	\mathbb{E}\left[ m_{\theta_0}(W) a(X) \left. \frac{\partial}{\partial t} \log dP_{\theta_0, \eta_t} \right|_{t=0} \right]  =0.
\]
Thus, under regularity conditions, the score of $P_t$ can be written as $h'\dot{\ell}_{\theta_0, P}+\dot l$,
where $\dot l$ belongs to
\[
	\dot{\mathbf{P}}_\eta = \left\{ \dot{l} \in L_0^2(P):  \mathbb{E} [ m_{\theta_0}(W) a(X) \dot{l}(X,W)]=0  \ \text{for any function} \ a \ \text{of} \ X \right\}.
\]
Conversely, under standard regularity conditions for conditional moment models, every element of $\dot{\mathbf{P}}_\eta$ can be generated by a differentiable nuisance path.
Under these conditions, the tangent set of $\mathbf{P}$ at $P$ is
\[
	T(P) = \left\{ h'\dot{\ell}_{\theta_0, P} + \dot{l}: h \in \mathbb{R}^p, \dot{l} \in \dot{\mathbf{P}}_\eta \right\}.
\]

\bibliographystyle{chicago}
\bibliography{specification.bib}

\end{document}